\documentclass{article}

\usepackage{apj}

\received{}
\accepted{}

\begin{document}

\title{Physical Parameters in the Hot Spots and Jets of  Compact Symmetric 
Objects}
\author{M. Perucho\altaffilmark{1} and J. M.$^{\underline{\mbox{a}}}$ Mart\'{\i}\altaffilmark{1}}
\altaffiltext{1}{Departamento de Astronom\'{\i}a y Astrof\'{\i}sica, Universidad 
de Valencia, 46100 Burjassot (Valencia), Spain. e-mail: manuel.perucho@uv.es, jose-maria.marti@uv.es}


\begin{abstract}
We present a model to determine the physical parameters of jets 
and hot spots of a sample of CSOs under very basic assumptions like synchrotron 
emission and minimum energy conditions. Based on this model we propose a simple 
evolutionary scenario for these sources assuming that they evolve in ram 
pressure equilibrium with the external medium and constant jet power. The 
parameters of our model are constrained from fits of observational data (radio 
luminosity, hot spot radius and hot spot advance speed) versus projected linear 
size. From these plots we conclude that CSOs evolve self-similarly and that 
their radio luminosity increases with linear size along the first kiloparsec. 
Assuming that the jets feeding CSOs are relativistic from both kinematical 
and thermodynamical points of view, we use the values of the pressure and particle 
number density within the hot spots to estimate the fluxes of momentum (thrust), 
energy, and particles of these relativistic jets. The mean jet power obtained in 
this way is within an order of magnitude that inferred for FRII sources, which is 
consistent with CSOs being the possible precursors of large doubles. The inferred 
flux of particles corresponds to, for a barionic jet, about a 10\% of the mass 
accreted by a black hole of $10^8 \, {\rm M_{\odot}}$ at the Eddington limit, 
pointing towards a very efficient conversion of accretion flow into ejection, 
or to a leptonic composition of jets.
We have considered three different models (namely Models, I, IIa, IIb). Model 
I assuming constant hot spot advance speed and increasing luminosity can be 
ruled out on the ground of its energy cost. However Models IIa and IIb seem to
describe limiting behaviours of sources evolving at constant advance speed and
decreasing luminosity (Model IIa) and decreasing hot spot advance speed and 
increasing luminosity (Model IIb). 
In all our models the slopes of the hot spot luminosity and advance speed with 
source linear size are governed by only one parameter, namely the external 
density gradient.
A short discussion on the validity of models II to describe the complete evolution
of powerful radio sources from their CSO phase is also included.      

\keywords{galaxies:active-galaxies:jets-galaxies:ISM-radio continuum:galaxies}
\end{abstract}

\section{Introduction} \label{s:intro}

  In the early eighties VLBI techniques allowed the discovery of compact, high 
luminosity radio sources with double structure and steep spectrum (Phillips \& 
Mutel 1980,1982). Some of them were found to have a core between the two outer 
components which were interpreted as lobes or hot-spots formed by a relativistic 
jet, and they were given the name of compact symmetric objects (CSOs) because of 
their double-sided emission and their small size (linear size lower than 1 kpc).

  The spectra of CSOs are steep with a peak at about 1 GHz, 
what makes them to belong to Gigahertz Peaked Spectrum Sources (GPSs, 
O'Dea et al.1991). If the peak is located around 100 Megahertz the source 
belongs to Compact Steep Spectrum Sources (CSSs, Fanti et al.1995). The 
smaller sources are more likely to have a GPS spectrum, while those with 
a projected linear size larger than one kpc (linear size between 1 and 20 
kpc) have a CSS spectrum. GPS and CSS sources include a variety of objects 
(O'Dea 1998), morphologically speaking, among which we find the double, 
symmetric ones: CSOs if their size is lower than 1 kpc (Wilkinson et al. 
1994), and Medium Size Symmetric Objects, MSOs, if their size does exceed 
1 kpc (Fanti et al. 1995). For a review about GPS and CSS sources see O'Dea 
(1998).

  The size of CSOs led radio astronomers to propose two opposed conjectures. One 
of them assumes a scenario where the external medium is so dense that the jet 
cannot break its way through it, so sources are old and confined (van Breugel 
et al. 1984), while the other one assumes that they are the young precursors of 
large symmetric sources like Faranoff-Riley type II galaxies (Phillips \& Mutel 
1982, Carvalho 1985, Mutel \& Phillips 1988). The former assumption is based on 
observations which show that some GPS sources are considerably optically 
reddened (O'Dea 1991), have distorted isophotes and disturbed optical 
morphologies, which indicate interaction with other galaxies or mergers. This 
can be interpreted as that the source has an abnormally dense medium, due to 
the gas falling onto the nucleus of the GPS from the companion. Under this 
assumption, sources can be confined by the external medium if it is dense 
enough (average number density of $10-100\, {\rm cm^{-3}}$), as it was shown by 
De Young (1991,93) through simulations of jet collisions with a dense, cloudy 
medium. Carvalho (1994,98) considers two scenarios, one where the NLR and ISM 
consist of a two-phase medium formed by a hot, tenuous one surrounding cold, 
dense clouds, with which the jet collides, mass loads and slows down, and 
another one where a uniform, dense external medium is assumed. This could 
result in the jet having to spend its life trapped within this medium and 
having ages of $10^{6}$ to $10^{7}$ years. On the other hand, it must be said 
that densities required for the jet to be confined imply huge masses for the 
innermost parsecs of the galaxy (De Young 1993).

  Recent measurements of component advance speeds for a few sources (Owsianik 
et al. 1998, Taylor et al. 2000), reveal that their speeds are better understood 
within the young source model, as they imply ages of no more than $10^{3}$ 
years. Theoretical evolutionary models have been proposed by Carvalho (1985), 
Readhead et al. (1996b), Fanti et al. (1995), Begelman (1996), O'Dea \& Baum 
(1997), Snellen et al. (2000), in which an attempt is made to establish a 
connection between CSOs, MSOs, and FRII. Simulations carried out by De Young 
(1993,97) also show that a jet evolving in a density gradient of a not very 
dense medium reproduces well those evolutionary steps.

  The study of CSOs is of interest because it will allow us to probe conditions
in the jet in the first kiloparsec of its evolution and the interaction with
the dense interstellar medium before it breaks through the intergalactic medium,
where jets have been extensively studied. Jets in CSOs are propagating through 
the NLR and ISM of AGNs, so this interaction is a good opportunity to get 
information about the central regions of AGNs, in particular about the central 
density and its gradient. Moreover, within the young source scenario, jets from
CSOs are in the earliest stages after their formation, allowing us to get 
information and constrain the conditions leading to the jet formation.

  In this paper, we obtain the basic physical parameters of jets and hot-spots 
of a sample of CSOs using very basic assumptions, in a similar way as Readhead 
et al. (1996a,b), i.e., synchrotron radiation theory, minimum energy assumption 
and ram pressure equilibrium with the external medium. We also propose a simple 
evolutionary scenario for them, based on observational data, through a 
theoretical model which gives the relevant magnitudes in the hot-spots as 
power laws of linear size. The model allows us to get some insight about the 
nature of CSOs and their environment, with the final aim of knowing whether 
these sources are related to large double radio sources. The criteria followed 
to obtain a sample of CSO and their data are explained in section 2. In section 
3, the theory used to get physical parameters for hot-spots out of their 
spectra is presented. In section 4 we use some basic assumptions to 
get information about the physical parameters of the jet.
Section 5 contains the evolutionary model proposed and 
comparison with previous models, and conclusions along 
with further comparison are presented in section 6. Finally, the relevant 
formulae used in the calculations of the physical magnitudes of hot-spots are 
presented in the Appendix. Throughout the paper we consider Hubble constant 
$H_{\rm 0}=100\, h \, {\rm km\,s^{-1}\,Mpc^{-1}}$, with normalised value 
$h=0.7$, and flat universe through a deceleration parameter $q_{\rm 0}=0.5$.

\section{A sample of CSOs} \label{s:sample}

  Sources have been selected from the GPS samples of Stanghellini et al. (1997),
Snellen et al. (1998, 2000) and Peck \& Taylor (2000). We have chosen those 
sources with double morphology already classified in the literature as CSOs and 
also those whose components can be safely interpreted as hot spots even though 
the central core has not been yet identified. The criteria we have followed are 
quite similar to those used by Peck \& Taylor(2000), i.e., detected core 
surrounded by double radio structure or double structure with edge brightening 
of both components; however, contrary to their criteria, we have included 
sources with an intensity ratio between both components greater than 10 in the 
frequency considered (see Table 1), relaxing this constraint to a 
value of 20 (in one source, 2128+048) and 11 for the rest. Anyway, sources 
possibly affected by orientation effects (beaming, spectra distortion) in a 
more evident way, like quasars and core-jet sources, have not been considered. 
The resulting sample is formed by 20 sources which are listed in 
Table 1 along with the data relevant for our study.

\section{Physical Parameters in the Hot Spots of CSOs}
\label{s:hotspots}

  Panels a) and b) of Fig. 1 display hot spot radius ($r_{\rm hs}$) 
and hot spot luminosity ($L_{\rm hs}$), respectively, versus projected source 
linear size ($LS$). These quantities are directly obtained from the 
corresponding measured (or modeled) angular sizes, flux densities in the 
optically thin part of the spectra and the formulae for cosmological distance 
(see Appendix for details). For those hot-spots with more than one component 
the radius was obtained as the one of the resulting total volume by adding the 
volumes of each component. One point per source is used by taking arithmetic 
mean values for the radius and radio-luminosity.

  Table 2 compiles the slopes for the corresponding linear 
log-log fits, the errors and the regression coefficients. A proportionality 
between hot spot radius and linear size is clearly observed. The hot spot 
luminosity seems to be independent of the source linear size, with only a weak 
tendency to grow with $LS$.
  
  In order to estimate internal physical parameters as the densities and 
energies of the ultrarelativistic particles in the hot spots, further 
assumptions should be made. According to the present understanding (see, for 
example, O'Dea 1998), the peak and inversion in the spectra of these sources is 
due to an absorption process which has been a matter of debate since the 
discovery of these objects. First, and most likely, synchrotron self absorption 
(SSA) may be the reason of the inversion, although Bicknell et al. (1997) and 
Kuncic et al. (1997), have proposed free-free absorption (FFA) and induced 
Compton scattering (ICS), respectively, as alternatives. Both latter models are 
successful in reproducing the decrease in peak frequency with linear size 
observed in GPS sources (O'Dea \& Baum 1997), but do not fit the data better 
than the SSA model. Also, Snellen et al. (2000) find evidence of SSA being the 
process of absorption producing the peak in GPS sources. Besides that, FFA and 
ICS do not allow us to extract information about the hot spot parameters, as 
absorption occurs in the surrounding medium of the hot-spots, by thermal 
electrons, whereas SSA occurs inside them.

  The problem with SSA model comes from its critical dependence on some
parameters (as an example, the magnetic energy density is proportional to the
tenth power of the peak frequency and the eighth power of the source angular
size) which makes it almost useless for our purposes. Having this in mind, we 
have relied on the minimum energy assumption, which states that the magnetic
field and particle energy distributions arrange in the most efficient way to 
produce the estimated synchrotron luminosity, as a conservative and consistent 
way to obtain information about the physical conditions in hot spots. As it is 
well known, the hypothesis of minimum energy leads almost to equipartition, in
which the energy of the particles is equal to that of the magnetic field. 
G\"uijosa \& Daly (1996) compared equipartition Doppler factors with those 
obtained assuming that X-ray emission comes from inverse Compton process for 
more than a hundred objects (including three radio galaxies also in our 
sample) concluding that they are actually near equipartition. 
Snellen et al. (2000) point out that sources must stay in equipartition if they 
are to grow self-similarly, as it seems to be the case (see 
Table 2. Finally, Table 3 in O'Dea (1998) compiles data from
Mutel et al. (1985) and Readhead et al. (1996a), and  compares magnetic field
estimates in the hot spots of several CSOs based on both minimum energy and 
SSA models. As both results are in rough agreement, the conclusion is that
sources undergo synchrotron self-absorption but are near equipartition.
Besides the minimum energy assumption, we also assume that there is no thermal
(barionic nor leptonic) component, so the number density of relativistic 
particles alone within the hot spots is estimated, and that each particle 
radiates its energy at the critical frequency, i.e., \emph{monochromaticity}.

  The calculation procedure for pressure ($P_{\rm hs}$) and number density of 
relativistic particles ($n_{\rm hs}$) is explained in the Appendix, and panels 
c) and d) of Fig. 1 represent their log-log plots versus projected 
source linear size for all the sources in our sample, along with the best linear
fit, assuming they all fulfill the minimum energy assumption. As in the case of 
panels a) and b) one point per source is plotted. We use volume weighted means 
of both magnitudes due to their intensive character. Slopes, errors, and 
regression coefficients of the corresponding fits are listed in 
Table 2.

  These plots and their fits may be interpreted as evolutionary tracks of the 
four magnitudes in terms of the distance to the origin, considering that this 
distance grows monotonically with time, as we will show in 
Sect.~(\ref{s:ssemcso}). Projection effects are surely a source of dispersion 
in the data, which on the other hand show good correlation. One way to test the 
influence of these projection effects is to use hot-spot radius, as it is not 
affected by projection, instead of linear size. Results for the fits are very 
similar (within error bars) to those in Table 2, so it can be 
stated that projection effects are not important as far as an evolutionary 
interpretation is concerned. We should keep in mind that we have removed from 
our sample those sources most likely pointing along the line of sight (quasars 
and core-jet sources).

  We can add to our series of data the recent measuremets of hot spot advance 
speeds (see Table 3). Owsianik \& Conway (1998) report a mean 
hot spot advance speed of $0.13h^{-1}c$ in 0710+439 whereas Owsianik et al. 
(1998b) conclude a speed of $0.10h^{-1}c$ in 0108+388. On the other hand, 
Taylor et al. (2000) give similar advance speeds for 0108+388 ($0.12h^{-1}c$) 
and 2352+495 ($0.16h^{-1}c$) while the speed they measure for 0710+439 is twice 
the one reported by Owsianik \& Conway (1998) ($0.26h^{-1}c$). Finally, 
Owsianik et al. (1998) derive an estimate for the hot spot advance speed of 
$0.13h^{-1}c$ for 2352+495, based on synchrotron ageing data from Readhead et 
al.(1996a) and measurements of the source size.
         
  The large difference of estimates in the case of 0710+439 can be attributed 
to a number of facts. On one hand, Taylor et al.'s (2000) measurements have 
been performed at a higher frequency which means that they have measured 
motions of a brighter and more compact working surface, which must be 
intrinsically faster than the lobe expansion. On the other hand, the velocity 
may have suffered a recent increase (Owsianik \& Conway 1998 data are concluded 
from five epochs from 1980 to 1993, whereas Taylor et al.'s 2000 measurements
are from three epochs from 1994 to 1999) as the authors point out. We should 
keep in mind that the jet is moving in a cloudy medium, the NLR or ISM, 
so measures of advance speed are conditioned by local environmental conditions.

 Finally, Taylor et al. (2000) detect motions for 1031+567, also included in our 
sample, for which an advance speed of $0.31h^{-1}c$ is inferred. However this 
speed is measured for one hot spot (component W1) and what could be a jet 
component (component E2) and, therefore, this speed may be overvalued.

  The results reported in the previous paragraphs concerning the hot spot 
advance speeds do not allow us to infer a definite behaviour of the hot 
spot advance speed with the distance to the source. However, excluding the
measurements of Taylor et al. (2000) on 0710+439 and 1031+567 for the above 
reasons, the remaining results are compatible with a constant expansion speed 
($v_{\rm hs} \propto LS^0$), that we shall assume as a reference in the 
evolution models developed in Sect.~(\ref{s:ssemcso}).

\section{Physical parameters in the jets of CSOs and the source energy budget}
\label{s:jets}

  Figure 2 shows a schematic representation of our model for CSOs in 
which the bright symmetric radio components are hot spots generated by the
impact of relativistic jets in the ambient medium. In the following we shall 
assume that the jets are relativistic from both kinematical and thermodynamical 
points of view, hence neglecting the effects of any thermal component. We can 
use the values of the pressure and particle number density within the hot spots 
to estimate the 
fluxes of momentum (thrust), energy, and particles of these relativistic jets. 
Under the previous hypothesis and assuming that hot spots advance at 
subrelativistic speeds, ram pressure equilibrium between the jet and hot spot 
leads to (Readhead et al. 1996a)
\begin{equation}
F_{\rm j} = P_{\rm hs}A_{\rm hs}, 
\end{equation}

\noindent
for the jet thrust $F_{\rm j}$, where $A_{\rm hs}$ stands for the hot spot cross 
section ($\simeq \pi r_{\rm hs}^2$). Taking mean values for $P_{\rm hs}$ and 
$r_{\rm hs}$ from our sample we get $F_{\rm j} \simeq(4.5\,\pm \, 3.3)
\,10^{34}$ dyn, where errors are calculated as average deviations from 
the mean.

  In a similar way, the flux of relativistic particles in the jet, $R_{\rm j}$, 
can be estimated from the total number of particles in the hot spot, 
$n_{\rm hs}V_{\rm hs}$ ($V_{\rm hs}$ is the hot spot volume, $\simeq 4 
\pi r_{\rm hs}^3/3$), and the
source lifetime, $\simeq v_{\rm hs}/LS$, where $v_{\rm hs}$ is the hot spot
advance speed. Assuming this speed constant and $\simeq 0.2c$, we can
write 
\begin{equation}
R_{\rm j} = n_{\rm hs}V_{\rm hs}v_{\rm hs}/LS \simeq (6.3\,\pm\,6.2)\,10^{48} 
{\rm e}^{+/-}{\rm s}^{-1}. 
\end{equation}
 
  Finally, a lower bound for the jet power, $Q_{\rm j}$, can be estimated 
considering that, in a relativistic jet from a thermodynamic point of view,
$Q_{\rm j} = (F_{\rm j}/v_{\rm j}) c^2$. Hence, for given $F_{\rm j}$ and
taking $v_{\rm j} = c$, we have 
\begin{equation}
Q_{\rm j, min} = F_{\rm j}c = P_{\rm hs} A_{\rm hs} c = 
 (1.3\,\pm\,1.0)\,10^{45} {\rm erg s}^{-1}.
\end{equation} 

  Let us point out that the values of the jet power and jet thrust derived
according to our model ($4.3-5.0\,10^{43}$ erg s$^{-1}$, $1.4-1.7\, 10^{33}$ 
dyn) are within a factor of 1.5 of those presented by Readhead et al. (1996a) 
for $h = 0.7$.

  Considering that the source spends the jet power in luminosity (basically, 
hot spot radio luminosity, $L_{\rm hs}$), advance ($Q_{\rm adv}$), and expansion 
of hot spots against the external medium ($Q_{\rm exp,hs}$), and that a fraction 
of the energy supplied is stored as internal energy of particles and magnetic 
fields in the hot spots ($\dot{U}_{\rm int,hs}$), we can write the following 
equation for the energy balance
\begin{equation}
Q_{\rm j} = L_{\rm hs} + \dot{U}_{\rm int,hs} + Q_{\rm adv} + Q_{\rm exp,hs} + 
Q_{\rm lobes},
\end{equation}

\noindent
where the term $Q_{\rm lobes}$ encompases the energy transferred to the lobes
(and cocoon) per unit time. Note that, in the previous equation, we have added 
the internal energy of the hot spots and the expansion work with respect to the 
work by Readhead et al. (1996a).

  The power invested by the source in advance and expansion of the hot spot and 
the variation of the internal energy in the hot spot per unit time can be
estimated as follows (assuming constant advance speed)
\begin{equation}
\dot{U}_{\rm int,hs} \approx P_{\rm hs}V_{\rm hs} 
\left( \frac{v_{\rm hs}}{LS}\right), 
\end{equation}
\begin{equation}
Q_{\rm adv} \approx P_{\rm hs} A_{\rm hs} v_{\rm hs},
\end{equation}
\begin{equation}
Q_{\rm exp,hs} \approx P_{\rm hs} \, 4 \pi r_{\rm hs}^2 v_{\rm hs} 
\left(\frac{r_{\rm hs}}{LS}\right),
\end{equation}

\noindent
where in the last expression we have used that the hot spot expansion speed is
$v_{\rm exp,hs} = v_{\rm hs} (r_{\rm hs}/LS)$, due to the self-similar evolution 
of the sources deduced from panel a) of Fig. 1, a result to be 
discussed in the next section. 

  Table 4 lists the average powers invested by the source in
their evolution for the values of the hot spot parameters derived in the
previous section. Despite the large uncertainties it is 
worthy to note that there seems to be some kind of equipartition between 
luminosity and expansion (source growth plus hot spot expansion) work per unit 
time. It has to be noted that percentages are obtained with respect to $Q_{\rm 
j, min}$. The remaining fraction, $45\%$, must be, at least in part, associated 
with power transferred to the lobes. Finally let us point that the increase of 
internal energy of the hot spot is a negligible fraction of the jet power.

\section{A Self-Similar Evolution Model for CSOs}
\label{s:ssemcso}

  In this section, we are going to construct an evolutionary model for CSOs 
based on the results presented in section \ref{s:hotspots}. The distance to 
the origin of the hot-spots will play the role of a time-like coordinate. The 
fits presented in that section will represent the evolution of the
corresponding physical quantity in a typical CSO helping us to contrain the
parameters of our model. 

 Our model is based on the assumption that the evolution of CSOs is dominated 
by the expansion of the hot spots as they propagate through the external 
medium. This conclusion is apparent after analising the source energy budget 
(see Table 4), as 33\% of $Q_{\rm j,min}$ is invested in 
expansion.

  We start by assuming that the linear size of the hot spot, $r_{\rm hs}$, grows 
with some power of time, i.e., $r_{\rm hs} \propto t^{\beta}$. We have chosen 
such a basic parameter because a value of $\beta$ can be easily deduced from the 
linear fits, as we shall see below. Our next assumption consists in considering 
a density decreasing external medium with $\rho_{\rm ext} \propto (LS)^{- 
\delta}$, with $\delta > 0$. In order to compare with the observational fits 
described in the previous section, we need to eliminate $t$ from our 
description. This is done through the velocity of advance of the hot spot, 
$v_{\rm hs}$, that fixes the dependence of the linear size of the source with 
the time. Considering that hot spots in CSOs are feed by relativistic jets but 
advance with significantly smaller speeds, the usual ram pressure equilibrium 
condition between the jet and the external medium leads to (Mart\'{\i} et al. 1997)
\begin{equation}
\label{eq:vhot}
v_{\rm hs} = \sqrt{\eta_{\rm R}  \frac{A_{\rm j}}{A_{\rm j,hs}} } v_{\rm j},
\end{equation}
where $\eta_{\rm R}$ is the ratio between the inertial density of the jet and 
that of the external medium ($\rho_{\rm ext}$), $A_{\rm j}$ and $A_{\rm j,hs}$ 
are, respectively, the cross-sectional area of the jet at the basis and at the 
hot spot, respectively, and $v_{\rm j}$ is the flow velocity in the jet. 
We can consider that $A_{\rm j,hs} \propto r_{\rm hs}^2$ and this is what we do 
in the next. Assuming that the jet injection conditions are constant with time 
we have

\begin{equation}
\left( \frac{dLS}{dt}  = \right) v_{\rm hs} \propto \left( \eta_{\rm R}  \frac{A_{\rm j}} 
{A_{\rm hs}} \right)^{1/2} v_{\rm j} \propto (LS)^{\delta/2} t^{-\beta},
\label{eq:dlsdt}
\end{equation}

\noindent
from which we derive the desired relation:

\begin{equation}
t \propto (LS)^{(1-\delta/2)/(1-\beta)}.
\label{eq:t}
\end{equation}

  Evolutionary tracks of sources that grow with time are obtained when the 
exponent in the later expression is positive, which means that both $\beta,
\delta/2 > 1$, or $\beta, \delta/2 < 1$. On the other hand, substituting this 
latter expression in eq.~(\ref{eq:dlsdt}) we find that 
\begin{equation}
v_{\rm hs} = (LS)^{(\delta/2-\beta)/(1-\beta)},
\end{equation}
from which we can conclude that the particular case $\beta = \delta/2$ 
(including the case $\beta = \delta/2 = 1$) leads to a constant hot spot advance 
speed and separates accelerating hot spot models ($\beta < 
\mbox{min}\{1,\delta/2\}$; $\beta > \mbox{max}\{1,\delta/2\}$) from decelerating ones ($\mbox{min}\{1,\delta/2\} < \beta < 
\mbox{max}\{1,\delta/2\}$).

  The hot spot radius in terms of the source linear size follows also from 
eq.~(\ref{eq:t})

\begin{equation}
r_{\rm hs} \propto (LS)^{\beta(1-\delta/2)/(1-\beta)}.
\end{equation}

  Self-similarity forces the exponent in this expression to be equal to 1
providing a relation between $\beta$ and the slope of the external density 
profile, $\delta$,
\begin{equation}  \label{eq:bedel}
\beta = \frac{2}{4-\delta}, 
\end{equation}
consistent with self-similar source evolution. Deduced from this expression is 
that $\beta \geq \delta/2$ which means that hot spots tend to decelerate within 
the first kpc if $\beta < 1$ or to accelerate if $\beta > 1$. We will discuss 
this result below. Note that our model allows for self-similar evolution tracks with
non-constant hot spot advance speeds, contrary to other models (e.g., 
Begelman 1996).

  The next equation in our model comes from the source energy balance. The 
energy injected by the jet is stored in the hot spots and lobes in the form
of relativistic particles, magnetic fields, and thermal material. Besides that, 
it provides the required energy for the source growth (hot spot expansion and
advance, lobe inflation). Finally, it is the ultimate source of luminosity.
Being CSO sources immersed in dense environments, a basic assumption is to 
consider that the work exerted by the hot spots against the external medium 
consumes a large part of jet power. This is, in fact, supported by the results
shown in previous section. Hence we assume 

\begin{equation}
(PdV)_{\rm hs,adv+exp} (\propto P_{\rm hs} r_{\rm hs}^2 LS) \propto t^{\gamma},
\end{equation}

\noindent
where the intermediate proportionality is, again, only valid for self-similar 
evolution. A value 1 for $\gamma$ would mean that the source adjusts its work 
per unit time to the jet power supply (that we consider as constant). 

  Finally, under the assumptions of minimum energy and monocromaticity (see 
Appendix), the luminosity of the hot spot, $L_{\rm hs}$, and the number density 
of relativistic particles, $n_{\rm hs}$, are found to follow the laws

\begin{equation}   
\label{eq:lhsi}
L_{\rm hs}\propto P_{\rm hs}^{7/4} r_{\rm hs}^3,
\end{equation}

\begin{equation}  
\label{eq:pn}
n_{\rm hs}\propto P_{\rm hs}^{5/4}.
\end{equation}

\subsection{Model I (3 parameters)}
\label{ss:modeli}

  The equations derived above can be manipulated to provide expressions for 
$v_{\rm hs}$, $r_{\rm hs}$, $L_{\rm hs}$, $P_{\rm hs}$, and $n_{\rm hs}$ in 
terms of source linear size, $LS$, 

\begin{equation}
v_{\rm hs} \propto (LS)^{(\delta/2-\beta)/(1-\beta)} 
\left( \propto (LS)^{\delta/2-1} \right)
\label{vhs_ls}
\end{equation}

\begin{equation}
r_{\rm hs} \propto (LS)^{\beta (1-\delta/2)/(1-\beta)}
\left( \propto LS \right)
\end{equation}

\begin{equation}
L_{\rm hs} \propto (LS)^{(7\gamma/4-\beta/2)(1-\delta/2)/(1-\beta)-7/4}
\left( \propto (LS)^{(7\gamma \, (2-\delta/2)-9)/4} \right)
\end{equation}

\begin{equation}
P_{\rm hs} \propto (LS)^{(\gamma-2\beta) (1-\delta/2)/(1-\beta)-1} 
\left( \propto (LS)^{\gamma(2-\delta/2)-3} \right)
\end{equation}

\begin{equation}
n_{\rm hs} \propto (LS)^{5/4 ( (\gamma-2\beta) (1-\delta/2)/(1-\beta)-1 ) }
\left( \propto (LS)^{5/4(\gamma(2-\delta/2)-3)} \right)
\label{nhs_ls}
\end{equation}

\noindent
where we have written in brackets the resulting expressions considering 
self-similarity, using the relation between $\beta$ and $\delta$ in eq. 
(\ref{eq:bedel}). Now, the first three relations (involving observable 
quantities) can be compared with the corresponding fits in section
(\ref{s:hotspots}) 
to obtain the values of the free parameters in our model, $\beta$, $\gamma$, and 
$\delta$. The comparison of the resulting power laws for $P_{\rm hs}$ and 
$n_{\rm hs}$ with their fits will provide a consistency test of the basic 
assumptions of our model. For constant hot spot advance speed the results are: 
$\beta=1.0\pm0.3$, $\delta=2.0\pm0.6$, $\gamma=1.5\pm0.3$, where errors are 
calculated from the obtained extreme values by changing the slopes of the fits 
within the given errors. The value of $\beta=1.0$ corresponds to a constant hot 
spot expansion speed. The value of $\delta=2$ is consistent with the external 
density profile in Begelman's (1996) model, for self-similar, constant growth 
sources. 

  The value obtained for $\gamma$ merits some discussion. In our present model,
the increase in luminosity inferred from the fits (and invoked by Snellen et
al. 2000 to explain the GPS luminosity function) does not need an external 
medium with constant density in the first kiloparsec (as concluded by Snellen
et al., 2000) but together with constant advance speed 
require that power invested by the hot spots in the advance 
and expansion work (see section \ref{s:jets}) grows with time as $t^{0.5}$. 
Taking into account that the expansion against the environment is a substantial 
fraction of the whole jet power supply, a value of $\gamma$ larger than 1 
implies that the expansion will eventually exhaust the source energy supply, 
producing a dramatic change in the source evolution (decrease in luminosity, 
deceleration of the hot spot advance) after the first kiloparsec. Recent 
calculations, in which we extend our study to MSO and FRII hot spots (Perucho
\& Mart\'{\i} 2001), show that radio luminosity in the hot spots (as well as
the expansion work) decrease in the long term. However, one should keep in mind 
that the trend of constant hot spot advance speed (and the luminosity growth 
with linear size) in the CSO phase are largely uncertain.

  The corresponding exponents for $P_{\rm hs}$ and $n_{\rm hs}$ ($-1.5\pm0.8$, 
$-1.9\pm 1.0$, respectively) are within the error bars of the fits presented
in section \ref{s:hotspots}, giving support to the minimum energy assumption 
considered in our model. 

\subsection{Model II (2 parameters)}
\label{ss:modelii}

  Model I has three free parameters which were fixed using the observational 
constraints. However, two of these constraints (namely hot spot luminosity 
versus source linear size, and hot spot advance speed versus source linear 
size) are poorly established. This is why we explore in this section two new 
models with only two free parameters by fixing $\gamma$ equal to 1. This is a 
reasonable choice as it expresses that the source self-adjusts the work per 
unit time to the (assumed constant) power jet supply. On the other hand, 
fixing one parameter allows us to liberate the models from one constraint 
allowing for the study of different evolutionary tracks. In particular we are 
going to study two models, IIa and IIb, although a continuity between them both 
is also possible, as discussed below.

  Making $\gamma = 1$ in eqs. (\ref{vhs_ls})-(\ref{nhs_ls}), we have
  
\begin{equation}
v_{\rm hs} \propto (LS)^{(\delta/2-\beta)/(1-\beta)}
\left( \propto (LS)^{\delta/2-1} \right)
\label{vhs_ls_ii}
\end{equation}

\begin{equation}
r_{\rm hs} \propto (LS)^{\beta (1-\delta/2)/(1-\beta)}
\left( \propto LS \right)
\end{equation}

\begin{equation}
L_{\rm hs} \propto (LS)^{(7/4-\beta/2)(1-\delta/2)/(1-\beta)-7/4}
\left( \propto (LS)^{(7\, (2-\delta/2)-9)/4} \right)
\end{equation}

\begin{equation}
P_{\rm hs} \propto (LS)^{(1-2\beta) (1-\delta/2)/(1-\beta)-1} 
\left( \propto (LS)^{-(\delta/2+1)} \right)
\end{equation}

\begin{equation}
n_{\rm hs} \propto (LS)^{5/4 ( (1-2\beta) (1-\delta/2)/(1-\beta)-1 )}
\left( \propto (LS)^{-5/4(\delta/2+1)} \right)
\label{nhs_ls_ii}
\end{equation}

Again results for self-similar evolution appear in brackets. Model IIa uses the 
fit for the $r_{\rm hs}$-$LS$ and constant speed assumption ($v_{\rm hs}$) to determine 
the values of $\beta$ and $\delta$. In Model IIb, the first condition is maintained 
(self-similarity) whereas the second is changed by the fit for radio-luminosity 
($L_{\rm hs}$-$LS$). 
The values of $\beta$ and $\delta$ for Models IIa and IIb as well as the exponents of the 
power laws for $v_{\rm hs}$, $r_{\rm hs}$, $L_{\rm hs}$, $P_{\rm hs}$, $n_{\rm 
hs}$ are listed in Table 5.

  Model IIa represents the self-similar evolution of sources with constant 
advance speed (what may be true, as indicated by the measurements of hot spot advance 
speeds, at least for the inner 100 parsecs). The decrease of density 
with linear size with an exponent of $-2$ is consistent with the values derived 
by other authors for larger scales (Fanti et al. 1995, Begelman 1996, De Young 
1993, 1997). Comparing with Model I, we see that constraining  $\gamma$ to 1.0 
leads to a decrease in luminosity while maintaining the hot spot expansion work.
The values of the exponents for $P_{\rm hs}$ and $n_{\rm hs}$ are in agreement
(whitin the respective errors bars) with those obtained in the fits. The energy 
required for the source to grow and expand at constant rate in the present model
without increasing the jet power supply (remember that now $\gamma = 1$) comes 
from a decrease in luminosity. In our model this decrease of the hot spot 
luminosity is produced by the fast reduction of pressure in the hot spot 
(caused by its fast expansion). However the required luminosity decrease 
($\propto (LS)^{-0.5}$) is quite far from the value derived for the 
$L_{\rm hs}$-$LS$ plot (despite its large error bar).

  Model IIb represents an extreme opposite case of Model IIa. Now, besides
self-similarity, we force the source to increase its luminosity at the rate
prescribed by the fit ($\propto (LS)^{0.3}$). The crucial parameter is, again
the density profile in the external medium that controls the expansion rate of 
the hot spot and the pressure decrease. The small external density gradient 
makes the source to decelerate its expansion rate maintaining a large pressure.
The values of the exponents of the hot spot pressure and density power laws are 
compatible (whitin the corresponding error bars) with those derived from the 
fits. The deceleration rate for the hot spot advance is large but plausible if 
one takes into account that the CSO hot spot advance speeds measured up to now 
(Owsianik \& Conway 1998, Owsianik et al. 1998b, Taylor et al. 2000) are all 
for small sources ($\le 100$ pc) which leaves lot of freedom for the hot spot 
advance speed profile in the first kpc. On the other hand the slowly decreasing 
external density profile ($\propto (LS)^{-1.1}$) is consistent with the 
structure of the ISM in ellipticals well fitted by King profiles with almost 
constant density galaxy cores 1 kpc wide. One model with constant external 
density ($\delta=0$) and self-similar expansion would have resulted in an 
increase of luminosity with distance to the source proportional to 
$(LS)^{1.25}$ and a decrease in hot spot pressure and advance speed as 
$(LS)^{-1}$. Such a large increase in luminosity is hardly compatible with the 
fit presented in section (\ref{s:hotspots}). Moreover, a density gradient like the 
one obtained in Model IIb allows for a smooth transition between the density in 
the inner core (which could be constant) and the gradient in outer regions, 
likely $-2$.

  Finally let us note that our hypothesis allow for a continuous transition 
between Models IIa and IIb by tuning the value of the exponent of the density 
power law between 1.1 and 2.0. In particular, the model with $\delta = 1.6$
fits very well the exponents of hot spot pressure (and relativistic particle
density) and predicts evolutive behaviours for $L_{\rm hs}$ and $v_{\rm hs}$ in 
reasonable agreement with the observable data ($L_{\rm hs}\propto LS^{-0.14}$ 
and $v_{\rm hs}\propto LS^{-0.2}$).

\section{Discussion}
\label{s:discussion}

  Results of the fits presented in section (\ref{s:hotspots}) show that sources 
evolve very close to self-similarity in the first kiloparsec of their lifes. 
This result agrees with what has been found by other groups. Snellen et al. 
(2000) calculate equipartition component sizes for a sample of GPS and CSS 
sources (Snellen et al. 1998a, Stanghellini et al. 1998, and Fanti et al. 1990) 
finding a proportionality with projected source overall size. Jeyakumar \& 
Saikia (2000) find self-similarity in a sample of GPS and CSS sources up to 20 
kpc. Concerning the dependence of radio luminosity with linear size, the fit
shown in section (\ref{s:hotspots}) points towards an increase of luminosity 
with linear size, as claimed by Snellen et al. (2000) for GPS sources. However, 
uncertainties are large and this dependence has to be confirmed by new CSO and 
GPS samples.

  As established in the Introduction, our study on CSOs offers an interesting 
link between fundamental parameters of the jet production process and the 
properties of large scale jets. It is interesting to note, on one hand, that the 
lower bound for the jet power is consistent (one order of magnitude larger) with 
the one inferred by Rawlings \& Saunders (1991) for FRII radio galaxies 
($10^{44}$ erg s$^{-1}$), supporting the idea of CSOs being the early phases of 
FRIIs. On the other hand, the flux of particles inferred in the jet is 
consistent with ejection rates of barionic plasma of the order $0.17\,M_\odot$ 
y$^{-1}$, implying a highly efficient conversion of accretion mass at the 
Eddington limit ($\dot{M_{\rm E}} \simeq 2.2\, M_{\rm \odot}$ y$^{-1}$, for a 
black hole of $10^{8}\, M_{\rm \odot}$) into ejection. The need for such a high 
efficiency could also point towards a leptonic composition of jets. Central densities
can be estimated using the ram pressure equilibrium assumption for those sources with 
have measured advance speeds, from the following equation equivalent to (\ref{eq:vhot})
$P_{\rm hs}= \rho_{\rm ext}\, v^2_{\rm hs}$. Results range from $1-10 {\rm cm^{-3}}$ for 
0108+108 which is close to the galactic nucleus, to $0.01-0.1 {\rm cm^{-3}}$ for 
2352+495 which is about 100 pc away.

  Our study concentrates in the evolution of sources within the first kpc 
assuming energy equipartition between particles and magnetic fields and hot spot 
advance in ram pressure equilibrium, extending the work of Readhead et al. 
(1996a,b) to a larger sample. In Readhead et al. (1996b) the authors construct 
an evolutionary model for CSOs based on the data of three sources (0108+388, 
0710+439, 2352+495) also in our sample. Comparing the properties of the two 
opposite hot spots in each source these authors deduce a value for the advance 
speed as a power law of external density approximately constant which fixes the 
remaining dependencies: $P_{\rm hs} \propto \rho_{\rm ext}^{1.00}$, $r_{\rm hs} 
\propto \rho_{\rm ext}^{-0.50}$. These results fit very well with those obtained 
in our Model IIa. Model IIb could be understood as complementary to Model IIa
and represent a first epoch in the early evolution of CSOs. It describes the 
evolution of a source in an external medium with a smooth density gradient, 
causing the decrease of the hot spot advance speed. During this first epoch, 
the luminosity of the source would increase. Then, the change in the external
density gradient (from -1.1 to -2.0) will stop the deceleration of the hot 
spots and would change the sign of the slope of luminosity, which now would
start to decrease (Model IIa). 

  To know whether CSOs evolve according to Model IIa or IIb (or a combination 
of both, IIb+IIa) needs fits of better quality. However, what seems clear is
that models with constant hot spot advance speed and increasing luminosity 
(i.e., Model I) can be ruled out on the ground of their energy costs.

  We can calculate the age of a source when it reaches 1 kpc (the edge of the
inner dense galactic core) according to Models IIa and IIb assuming an initial 
speed (let say at 10 pc) of 0.2 c (as suggested by recent measurements). In the
case of Model IIa this age is of $3.3\,10^{4}$ y, whereas in the case of Model 
IIb the age is about one order of magnitude larger (i.e., $\simeq 1.1\,10^{5}$ 
y). In this last case, the source would reach this size with a speed of 0.02 c.
If hot spots advance speeds remain constant after 1 kpc (consistent with a
density gradient of slope $-2.0$, commonly obtained in fits for large scale
sources; see below) then the age of a source of size 100 kpc would be of the
order $1.6\,10^{7}$ y. This result supports CSOs as precursors of large FRII 
radio sources.

  Since the work of Carvalho (1985) considering the idea of compact doubles 
being the origin of extended classical doubles, several attempts have been made 
to describe the evolution of CSOs to large FRII sources. Fanti et al. (1995) 
discussed a possible evolutionary scenario based on the distribution of sizes 
of a sample of CSS sources of {\it medium} size ($< 15$ kpc), assuming 
equipartition and hot spot advance driven by ram pressure equilibrium. Their 
model supports the young nature of MSOs predicting a decrease in radio 
luminosity by a factor of ten as they evolve into more extended sources and that 
external density changes as $(LS)^{-2}$ after the first half kiloparsec.

  Begelman's (1996) model predicts an expansion velocity depending only weakly 
on source size and a evolution of luminosity proportional to $\approx 
(LS)^{-0.5}$ for ambient density gradients ranging from $(LS)^{-(1.5-2.0)}$. It 
accounts for the source statistics and assumes Begelman \& Cioffi's (1989) model 
for the evolution of cocoons surrounding powerful extragalactic radio sources. 
This means constant jet power, hot spot advance driven by ram pressure 
equilibrium, internal hot spot conditions near equipartition, and that internal 
pressure in the hot-spots is equal to that in the cocoon multiplied by a 
constant factor, 
condition which turns to be equivalent to self-similarity. Snellen et al. (2000) 
explain the GPS luminosity function with a self-similar model, assuming again 
constant jet power. The model predicts a change in the slope of the radio 
luminosity after the first kpc ($\propto (LS)^{2/3}$ in the inner region; 
decreasing at larger distances), governed by an external density King profile 
falling with $(LS)^{-1.5}$ outside the 1 kpc core.  

  Model IIa predicts an external density profile in agreement with those 
inferred in the long term evolution models just discussed. However, it 
leads to a decrease of the hot spot pressure ($\propto (LS)^{-2}$) too large.
Readhead et al. (1996b) compare their data for CSOs with more extended 
sources (quasars from Bridle et al. 1994) and obtain a best fit for pressure 
$P_{\rm hs} \propto (LS)^{-4/3}$, which is in agreement with our Model IIb.
However, Model IIb produces a flat ($\propto (LS)^{-1.1}$) external density
gradient and an unwelcome increase in radio luminosity. The conclusion is that 
neither Model IIa nor IIb can be directly applied to describe the complete
evolution of powerful radio sources from their CSO phase. In Perucho \& Mart\'{\i} 
(2001) we have plotted the same physical magnitudes than here versus projected linear 
size for sources that range from CSOs to FRIIs and
the most remarkable fact is the almost constant slope found for pressure evolution.
We try to reconcile the change of the slopes found for 
external density and luminosity with this behaviour of pressure by considering a time 
dependent (decreasing) jet power, in agreement with 
the jet powers derived for CSOs and FRIIs (a factor of 20 smaller the latter). In 
that paper, we have used the samples of MSOs given by Fanti et al. (1985) and 
FRIIs by Hardcastle et al. (1998) and estimated the relevant physical magnitudes 
in their hot spots as we have done for CSOs. The fit for the plot of hot-spot 
radius versus projected linear size shows the loss of self-similarity from 10 
kpc on, a result which is consistent with that of Jeyakumar \& Saikia (2000). 
Concerning radio luminosity, a clear break in the slope at 1 kpc is apparent, 
feature predicted by Snellen et al. (2000) for GPSs. 

\section{Conclusions}
\label{s:concl}

  In this paper, we present a model to determine the physical parameters of 
jets and hot-spots of a sample of CSOs under very basic assumptions like 
synchrotron emission and minimum energy conditions. Based on this model 
we propose a simple evolutionary scenario for these sources assuming that 
they evolve in ram pressure equilibrium with the external medium and constant
jet power. The parameters of our model are constrained from fits of 
observational data (radio luminosity, hot spot radius, and hot spot advance 
speeds) and hot spot pressure versus projected linear size. From these plots we
conclude that CSOs evolve self-similarly (Jeyakumar \& Saikia 2000) and that 
their radio luminosity increases with linear size (Snellen et al. 2000) along 
the first kiloparsec. 

  Assuming that the jets feeding CSOs are relativistic from both kinematical 
and thermodynamical points of view, hence neglecting the effects of any thermal 
component, we use the values of the pressure and particle number density within 
the hot spots to estimate the fluxes of momentum (thrust), energy, and particles 
of these relativistic jets. We further assume that hot spots advance at
subrelativistic speeds and that there is ram pressure equilibrium between the 
jet and hot spot. The mean jet power obtained in this way is, within an order of 
magnitude, that given by Rawlings \& Saunders (1991) for FRII sources, which is 
consistent with them being the possible precursors of large doubles. The 
inferred flux of particles corresponds to, for a barionic jet, about a 10\% of 
the mass accreted by a black hole of $10^8 \, {\rm M_{\odot}}$ at the Eddington 
limit, pointing towards a very efficient conversion of accretion flow into 
ejection, or to a leptonic composition of jets.

  We have considered three different models (namely Models, I, IIa, IIb). Model 
I assuming constant hot spot advance speed and increasing luminosity can be 
ruled out on the ground of its energy cost. However Models IIa and IIb seem to
describe limiting behaviours of sources evolving at constant advance speed and
decreasing luminosity (Model IIa) and decreasing hot spot advance speed and 
increasing luminosity (Model IIb). However, in order to know whether CSOs evolve 
according to Model IIa or IIb (or a combination of both, IIb+IIa) we need fits of 
better quality and more determinations of the hot spot advance speeds and radio-luminosity. 
In all our models the slopes of the hot spot luminosity and advance speed with 
source linear size are governed by only one parameter, namely the external 
density gradient.
  
  Terminal speeds obtained for Model IIb, in which we find a negative slope 
for the hot spot advance speed, are consistent with advance speeds inferred for 
large sources like Cygnus A (Readhead et al. 1996b). This fact, together
with the ages estimated from that model and the recent measures of advance 
speed of CSOs (Owsianik et al. 1998, Taylor et al. 2000) support the young
scenario for CSOs. Moreover, central densities estimated in section 
(\ref{s:discussion}) using ram pressure equilibrium assumption are low enough 
to allow jets with the calculated kinetic powers to escape (De Young 1993).  
External density profile in Model IIa is consistent with that given for large 
sources ($-2.0$), while Model IIb gives a smoother 
profile as corresponds to a King profile in the inner kiloparsec.

  Although Models II seem to describe in a very elegant way the evolution 
of CSOs within the first kpc, preliminary results show that neither Model IIa 
nor IIb can be directly applied to describe the complete evolution of powerful 
radio sources from their CSO phase. In Perucho \& Mart\'{\i} (2001) we try to 
reconcile the change of the slopes of external density and luminosity with the 
behaviour of pressure (see discussion section) by considering a time dependent 
(decreasing) jet power.

\begin{acknowledgements}

  We thank A. Peck and G.B. Taylor for the data they provided us. We thank D.J. 
Saikia for his interest in our work and supply of information about his papers, 
which were very useful for us. This research was supported by Spanish 
Direcci\'on General de Investigaci\'on Cient\'{\i}fica y T\'ecnica (grants 
DGES-1432). 
  
\end{acknowledgements}
  
\section*{Appendix: Obtaining basic physical parameters from observational data}

\subsection*{1. Intrinsic luminosities and sizes}

We obtain the required parameters by using observational data in a simple way.
The first step is to obtain the luminosity distance to the source, in terms of 
redshift and the assumed cosmological model,

\begin{equation}
D_{\rm L}=\frac {c\,z}{H_{\rm 0}}\left( \frac {1+\sqrt{1+2q_{\rm 0}z}+z} 
{1+\sqrt{1+2q_{\rm 0}z}+q_{\rm 0}z} \right)
\end{equation}

Angular distance, used to obtain intrinsic linear distances, is defined as

\begin{equation}
\label{eq:dth}
D_{\rm \theta}=\frac{D_{\rm L}} {(1+z)^2}
\end{equation}

Intrinsic linear distances (like the source linear size, $LS$, and hot spot
radius, $r_{\rm hs}$) are obtained from the source angular distance and the
corresponding angular size of the object

\begin{equation}
LS = D_{\rm \theta}\,\theta_{\rm T}/2 \quad , \quad  
r_{\rm hs} = D_{\rm \theta}\,\theta_{\rm hs} 
\end{equation}

\noindent
with $\theta_{\rm T}$ being the total source angular size, and $\theta_{\rm hs}$ 
the angular size of the hot spot, given in Table 1. The 
intrinsic total radio luminosity of the hot spots can be obtained in terms of 
the observed flux density, $S_{\rm hs}$, and the luminosity distance according
to

\begin{equation}
L_{\rm hs} = 4\pi\,D_{\rm L}^2 S_{\rm hs},
\end{equation}

\noindent
where $S_{\rm hs}$ corresponds to the total flux in the frequency range
$10^{7} - 10^{11}$ Hz ($\nu_{\rm min}$, $\nu_{\rm max}$, in the following)

\begin{equation}
\label{eq:lhs}
S_{\rm hs} = \int_{\rm \nu_{\rm min}}^{\rm \nu_{\rm max}} C \, (\nu)^{-\alpha} 
d\nu.
\end{equation}

The constant for the spectrum in the observer reference frame can be obtained 
from the flux density at a given frequency ($\nu_{\rm 0}$) and the spectral 
index, having in mind that the synchrotron spectra in the optically thin limit
follows a power law

\begin{equation}
C = S_{\rm \nu_{\rm 0}} \, (\nu_{\rm 0})^{\alpha}.
\end{equation}

Hence, in terms of known variables, the total intrinsic radio luminosity is 
written as follows

\begin{equation}
L_{\rm hs} = S_{\rm \nu_{\rm 0}}\,4\pi\,D_{\rm L}^2\, \nu_{\rm 0}^{\alpha} \,
\frac{(\nu_{\rm max})^{1-\alpha} - (\nu_{\rm min})^{1-\alpha}}{1-\alpha} \quad 
(\alpha \neq 1)
\end{equation}

\begin{equation}
L_{\rm hs} = S_{\rm \nu_{\rm 0}}\,4\pi\,D_{\rm L}^2\, \nu_{\rm 0}^{\alpha} \,
\ln \left(\frac{\nu_{\rm max}}{\nu_{\rm min}}\right)  \quad   (\alpha = 1).
\end{equation}

\subsection*{2. Minimum Energy Assumption}

  Once obtained intrinsic sizes and luminosities from observations, the next 
step is to use them to constrain physical parameters (like pressure, magnetic
field strength, and relativistic particle density) in the hot spots. Our model
is based on the minimum energy assumption according to which the magnetic field 
has such a value that total energy of the object is the minimum necessary so as 
to produce the observed luminosity. As it is well known, this assumption leads
almost to the equipartition of energy between particles and magnetic field.

 The total internal energy of the system can be written in terms of the magnetic 
field strength, $B$. First, the energy density of the relativistic particles 
\begin{equation}
u_{p}=\int^{E_{\rm max}}_{E_{\rm min}} n(E)\, E\, dE
\label{eq:up}
\end{equation}

\noindent
(where $E$ is the energy of particles and $n(E)$, the number density at the
corresponding energy) can be estimated assuming monochromatic emission. 
According to this, any electron (of energy $E$) radiates only at its critical 
frequency, given by

\begin{equation}    
\label{eq:nur}
\nu_{\rm c} = C_{1}\,B\,E^2,
\end{equation}

\noindent
where $C_{1}$ is a constant ($\simeq 6.3 \, 10^{18}$ in cgs units). The
monochromatic emission assumption allows to change the energy integral in 
Eq.(\ref{eq:up}) by an integral of the intrinsic emitted flux in the
corresponding range of critical frequencies. At the end, the total internal 
particle energy in the hot spots, $U_{\rm p}$, is (see, e.g., Moffet 1975)
\begin{equation}
U_{\rm e}=A  L_{\rm hs}  B^{-3/2},
\end{equation}

\noindent
where $A$ depends on the spectral index and the frequency range:

\begin{equation}
A = \frac{C_{\rm 1}^{1/2}}{C_{\rm 3}}  \frac{2-2\alpha}{1-2\alpha}
\frac{\nu_{\rm max}^{(1/2)-\alpha}-\nu_{\rm min}^{(1/2)-\alpha}}
{\nu_{\rm max}^{1-\alpha}-\nu_{\rm min}^{1-\alpha}},
\end{equation}

\noindent
($C_{\rm 3}\simeq 2.4\, 10^{-3}$ in cgs units). Limits for the 
radio-emission frequencies are taken to be $10^{7}$ and $10^{11}$ Hz. $A$
varies within a factor of six ($3.34\, 10^{7}$, $2.2\, 10^{8}$) for
extreme values of the spectral index, $\alpha$ (0.75, 1.5, respectively).

  Then, the expression for the total internal energy in the hot-spot is

\begin{equation}  
\label{eq:enmin}
U_{\rm tot} = U_{\rm p} + U_{\rm B} = A  L_{\rm hs}  B^{-3/2} + V_{\rm hs}  
\frac {B^{2}}{8\pi},
\end{equation}

\noindent
where $V_{\rm hs}$ is the volume of the hot spot (assumed spherical) and 
$B^{2}/8\pi$ is the magnetic field energy density. The magnetic field which 
gives the minimum energy for the component comes directly from minimising 
equation (\ref{eq:enmin}), leaving as constant the intrinsic luminosity of the
hot spot, $L_{\rm hs}$,

\begin{equation}       
\label{eq:bmin}
B_{\rm min}= \left( \frac {6\pi A  L_{\rm hs}}{V_{\rm hs}} \right)^{2/7},
\end{equation}

Hence, the energies associated to magnetic field and relativistic particles are,
finally,

\begin{equation}
u_{\rm B}=\frac {B_{\rm min}^{2}}{8\pi},
\end{equation}

\noindent
and 

\begin{equation}
u_{\rm p} = (4/3)  u_{\rm B}.
\end{equation}

  Pressure at the hot spots has two contributions
\begin{equation}      
\label{eq:pmin}
P_{\rm hs} = (1/3) u_{\rm p} + (1/3) u_{\rm B} = (7/9) u_{\rm B}.
\end{equation}
The number density of relativistic particles follows from the monochromatic
emission assumption:
\begin{equation}
\label{eq:nel}
n_{\rm hs} = \frac{L_{\rm hs}C_{\rm 1}}{C_{\rm 3}B_{\rm min} V_{\rm hs}}  
             \frac{2\alpha-2}{2\alpha}
             \frac{\nu_{\rm max}^{-\alpha}-\nu_{\rm min}^{-\alpha}}
             {\nu_{\rm max}^{1-\alpha}
             -\nu_{\rm min}^{1-\alpha}}.
\end{equation}

\newpage

\begin{figure*}
\plotone{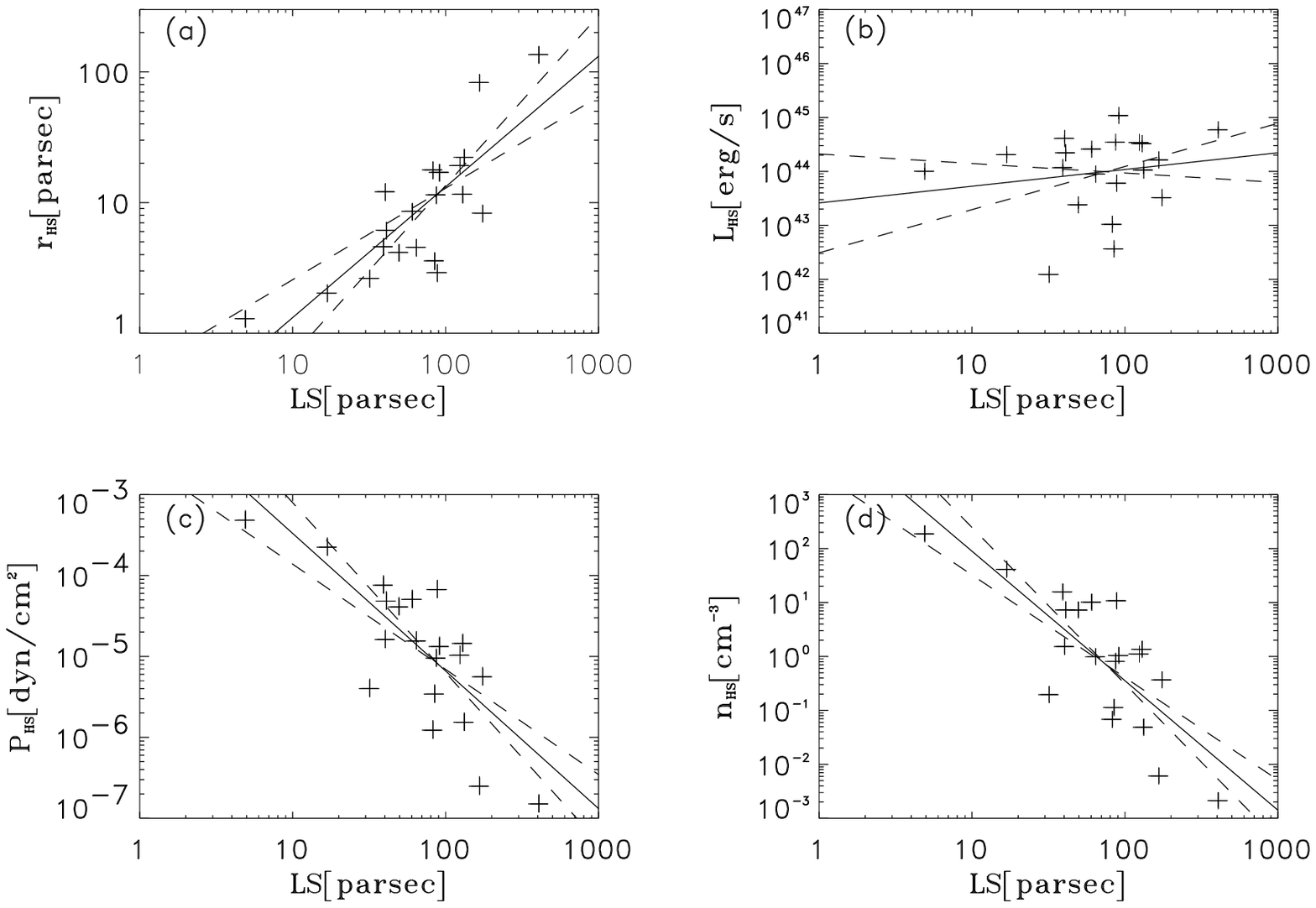}
\figcaption{Log-log plots of radius (panel a), radio-luminosity (panel b), pressure (panel c) and 
density (panel d) of hot-spots versus projected linear size. One point per CSO 
is plotted (see text). Error bars correspond to 15\% in angular sizes for 
radius, pressure and density and 15\% in measured radio-flux at the given 
frequency for radio-luminosity, and they are just indicative. Dashed lines give
account of limiting slopes.} 
\label{rls}
\end{figure*} 

\begin{figure*}
\plotone{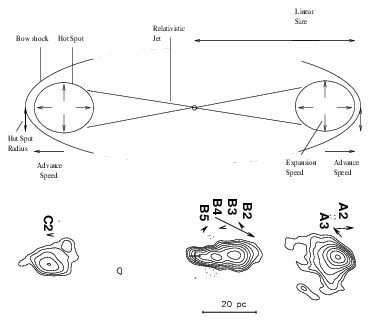}
\figcaption{Schematic view of a CSO with a radio-image of 0710+439 (Taylor et al. 2000).} 
\label{model}
\end{figure*}

\begin{deluxetable}{c c c c c c c c}
\tablefontsize{\scriptsize}
\label{tab:dat}
\tablecolumns{8}
\tablewidth{0pc}
\tablehead{
Source&$\theta$(mas)&$\theta_{\rm T}$(as)&Spectral&z&$\nu$(GHz)&
$S_{\rm \nu}$(Jy)&Refs.\nl
&&&index($\alpha$)&&&&}

\startdata
0108+388S& 0.821&   0.006&   0.900&   0.669&   15.36&   0.118&1,2,3\nl
0108+388N&   0.586&   0.006&   0.900&   0.669&   15.36&   0.172&1,2,3\nl

0404+768E&   45.0&   0.150&   0.501&   0.599&   1.70&   0.429& 1,2,7\nl
0404+768W&   54.0&   0.150&   0.501&   0.599&   1.70&   4.181& 1,2,7\nl  

0500+019N&   5.17&   0.015&   0.900&   0.583&  8.30&   1.25& 1,5,10\nl
0500+019S&   3.49&   0.015&   0.900&   0.583&   8.30&   0.110& 1,5,10\nl

0710+439N&   0.950&   0.025&   0.600&   0.518&   8.55&   0.330&1,3,4\nl
0710+439S&   2.16&   0.025&   0.600&   0.518&   8.55&   0.110&1,3,4\nl

0941-080N&   7.637&   0.050&   1.01&   0.228&   8.30&   0.080&1,5\nl
0941-080S&   12.7&   0.050&   1.01&   0.228&   8.30&   0.130&1,5\nl

1031+5670W&  1.047&  0.036&   1.10&   0.460&   15.3&   0.080&17\nl
1031+5670E&  1.296&  0.036&   0.80&   0.460&   15.3&   0.065&17\nl

1111+1955N&  2.800&  0.020&  1.50&   0.299&  8.40&  0.126& 17,18,19,20\nl
&1.070&&&&&&\nl
1111+1955S&  1.370&  0.020&  1.50&   0.299&  8.40&  0.090& 17,18,19,20\nl

1117+146N&   4.40&   0.080&   0.800&   0.362&   22.9&   0.050& 1,11\nl   
1117+146S&   2.90&   0.080&   0.800&   0.362&   22.9&   0.100& 1,11\nl

1323+321N&   9.83&   0.060&   0.600&   0.369&   8.55&   0.700&1,4\nl   
1323+321S&  10.28&   0.060&   0.600&   0.369&   8.55&   0.380&1,4\nl

1358+624N&   27.0&   0.070&   0.700&   0.431&   1.663&   1.152&1,4\nl
1358+624S&   40.2&   0.070&   0.700&   0.431&   1.663&   2.601&1,4\nl

1404+286N&   0.990&   0.007&   1.60&   0.077&   8.55&   1.67&1,4\nl   
&1.19&&&&&&\nl
1404+286S&   2.14&   0.007&   1.60&   0.077&   8.55&   0.140&1,4\nl   
&0.380&&&&&&\nl

1414+455N&   3.20&   0.034&   1.62&   0.190&  8.40&   0.042& 17,18,19,20\nl
1414+455S&   2.30&   0.034&   1.52&   0.190&  8.40&   0.034& 17,18,19,20\nl
& 1.15&&&&&&\nl

1607+268N&   3.78&   0.050&   1.20&   0.473&   5.00&   0.840& 1,13\nl   
&6.66&&&&&&\nl
1607+268S&   6.48&   0.050&   1.20&   0.473&   5.00&   0.740&1,13\nl  
&6.66&&&&&&\nl

1732+094N&   1.947&   0.015&   1.10&   0.610&   5.00&   0.480&1,15\nl  
1732+094S&   2.48&   0.015&   1.10&   0.610&   5.00&   0.285&1,15\nl

1816+3457N&  4.46&   0.035&   1.92&   0.245&  8.40&   0.028&17,18,19,20\nl
1816+3457S&  4.57&   0.035&   1.85&   0.245&  8.40&   0.074&17,18,19,20\nl
&1.67&&&&&&\nl  
&3.73&&&&&&\nl

1946+704N&   1.46&   0.036&   0.640&   0.101&   14.9&   0.122& 8,9\nl   
&2.42&&&&&&\nl
1946+704S&   3.27&   0.036&   0.640&   0.101&   5.00&   0.019& 8,9\nl

2008-068N&   2.74&   0.030&   0.800&   0.750&   5.00&   1.01& 1,15\nl   
2008-068S&   4.68&   0.030&   0.800&   0.750&   5.00&   0.112& 1,15\nl

2050+364W&   3.06&   0.060&   0.900&   0.354&   5.00&   2.11& 1,12,13,14\nl  
2050+364E&   5.22&   0.060&   0.900&   0.354&   5.00&   2.89& 1,12,13,14\nl
&3.60&&&&&&\nl
&4.50&&&&&&\nl

2128+048N&   4.62&   0.030&   0.800&   0.990&   8.30&   1.21& 1,5,10\nl   
&5.80&&&&&&\nl
2128+048S&   3.82&   0.030&   0.800&   0.990&   8.30&   0.060& 1,5,10\nl

2352+495N&   1.10&   0.050&   0.501&   0.237&   5.00&   0.080& 1,2,16\nl   
2352+495S&   2.60&  0.050&   0.501&   0.237&   5.00&   0.040& 1,2,16\nl
\enddata
\tablecomments{
Data in the columns: (1) B1950.0 coordinates (N means northern hot-spot, S 
southern, etc.); (2) angular size ($\theta$) of the hot spots (or 
subcomponents); (3) total angular size of the radio source, ($\theta_{\rm T}$); 
(4) spectral index, $\alpha$, of the optically thin part of the spectrum; 
(5) source redshifts, $z$; (6) and (7) frequency ($\nu$) and flux density 
($S_{\rm \nu}$) at that frequency of the optically thin part of the spectrum 
for the whole hot spot (adding the fluxes of subcomponents if necessary); 
(8) references from which the data of each source have been taken. In those 
cases in which no spectral index for each hot spot were available, we have used 
that of the whole source. Angular sizes were taken for the highest frequency 
available in order to eliminate the contribution of the diffuse component. The 
angular sizes of components have been calculated by multiplying the geometric 
mean of the FWHM Gaussian axes by a factor 1.8, following Readhead et al. 
(1996a), in order to have spherical hot spots.}
\tablerefs{
(1)  Stanghellini et al. 1998, A\&AS, 131, 303; (2)  Taylor et al. 1996, ApJ, 463, 
95; (3)  Pearson et al. 1988, ApJ, 328, 114; (4)  Fey et al. 1996, A\&AS, 105, 
299; (5)  Dallacasa et al. 1998, A\&AS, 129, 219; (6)  Xu et al. 1995, A\&AS,
99, 297; (7)  Dallacasa et al. 1995, A\&AS, 295, 27; (8)  Snellen et al. 1998a, 
A\&AS, 131, 435; (9)  Snellen et al., astro-ph/0002129; (10) Stanghellini 
et al. 1997, A\&A, 325, 943; (11) Bondi et al. 1998, MNRAS, 297, 559; (12) 
Phillips \& Mutel 1981, ApJ, 244, 19; (13) Mutel et al. 1985, ApJ, 290, 86; 
(14) Mutel et al. 1986, ApJ, 307, 472; (15) Stanghellini et al. 1999, A\&AS, 134, 
309; (16) Readhead et al. 1996a, ApJ, 460, 612; (17) Taylor et al., 
astro-ph/0005209; (18) Peck \& Taylor 2000, ApJ, 534, 104; (19) Peck et al. 2000, 
ApJ, 534, 90; (20) A.Peck 2000, private communication. (c):detected core.}
\end{deluxetable}

\begin{deluxetable}{lccc}
\tablefontsize{\footnotesize}
\tablecolumns{4}
\tablewidth{0pc}
\tablecaption{Best fits of physical parameters in hot-spots in terms of source 
linear size.} 
\label{tab:fits} 
\tablehead{& $\alpha$ & $\varepsilon$ & $r$}    
\startdata
$r_{\rm hs}$ & $ 1.0$   & $0.3$         & $ 0.80$ \nl
$L_{\rm hs}$ & $ 0.3$   & $0.5$         & $ 0.17$ \nl
$P_{\rm hs}$ & $-1.7$   & $0.4$         & $-0.79$ \nl
$n_{\rm hs}$ & $-2.4$   & $0.5$         & $-0.78$ \nl
\enddata
\tablecomments{ 
$\alpha$ is the slope of the corresponding log-log best fit, $\varepsilon$ is 
the error of that fit, and $r$ is the regression coefficient.}
\end{deluxetable}

\begin{deluxetable}{c c c c}
\tablefontsize{\footnotesize}
\tablecolumns{4}
\tablewidth{0pc}
\tablecaption{Hot-spot advance speed values} \label{tab:vel}
\tablehead{Source& Owsianik et al.& Taylor et al.& $LS$ (pc)}
\startdata
0108+388 & $0.098\, h^{-1}\,c$    & $0.12\, h^{-1}\,c$ & 17\nl
0710+439 & $0.13\, h^{-1}\,c$     & $0.26\, h^{-1}\,c$ & 64\nl
1031+567 & -                      & $0.31\, h^{-1}\,c$ \tablenotemark{a} & 88\nl
2352+495 & $0.13\, h^{-1}\,c$ \tablenotemark{b} & $0.16\, h^{-1}\,c$ & 85\nl
\enddata
\tablenotetext{a}{ \footnotesize{Speed measured for one hot spot and, possibly, a jet 
component.}} 
\tablenotetext{b}{\footnotesize{Calculated from synchrotron ageing data from Readhead et al. 
(1996a).}}
\end{deluxetable}

\begin{deluxetable}{lcccc}
\tablefontsize{\footnotesize}
\tablecolumns{5}
\tablewidth{0pc} 
\tablecaption{Powers invested by the jets in their evolution.}   
\label{tab:powers}

\tablehead{ & $L_{\rm hs}$ & $\dot{U}_{\rm int,hs}$ & $Q_{\rm adv}$ & 
$Q_{\rm exp,hs}$ }
\startdata
Power      & $(2.2 \pm 2.1)\,10^{44}$ & $(8.3 \pm 7.8)\,10^{43}$ & 
$(2.5 \pm 1.9)\,10^{44}$ & $(1.9 \pm 1.8)\,10^{44}$ \nl
Fraction   & $0.16\pm 0.16$ & $0.06\pm 0.06$ & $0.19$ & $0.14\pm0.14$ \nl
\enddata
\tablecomments{Values are in erg s$^{-1}$. Fractions are over $Q_{\rm j,min}$. 
Errors are calculated as for jet parameters.}
\end{deluxetable}

\begin{deluxetable}{cccccccc}
\tablefontsize{\footnotesize}
\tablecolumns{8}
\tablewidth{0pc}
\tablecaption{Exponents of evolution for Models IIa and IIb}
\label{tab:exps}
\tablehead{
Model & $\beta$ & $\delta$ & $r_{\rm hs}$ & $L_{\rm hs}$ & $v_{\rm hs}$ & 
$P_{\rm hs}$ & $n_{\rm hs}$}
\startdata
IIa & $1.0\pm0.3$ & $2.0\pm0.6$ & 1.0 & $-0.50\pm0.15$ & 0.0 & $-2.0\pm0.6$ & 
$-2.50\pm0.75$ \nl
IIb & $0.7\pm0.3$ & $1.1\pm1.0$ & 1.0 & 0.3 & $-0.5\pm0.4$ & $-1.5\pm0.8$ & 
$-1.9\pm1.0$ \nl
\enddata
\tablecomments{Exponents of the linear size power law fits of the physical 
parameters in the CSO hot spots for Models IIa and IIb. Errors, as before, were 
obtained from the extreme values given by the errors obtained for the slopes in 
the fits.}
\end{deluxetable}

\end{document}